\documentclass[
    ,final            
  ,dvips]
  {aipproc}

\layoutstyle{8x11double}

\begin{document}

\title[]{Spin-dependent Proximity Effects in d-wave Superconductor/Half-metal Heterostructures}


\author{Nobukatsu Yoshida}
{address={Department of Physics, University of Tokyo, Hongo,Tokyo 113-0033, Japan}}

\author{Mikael Fogelstr\"om}
{address={Applied Quantum Physics, MC2, Chalmers University of Technology, S-412 96 G\"oteborg, Sweden}}

\begin{abstract}
We report on mutual proximity effects in $d$-wave superconductor/half-metal heterostructures
which correspond to systems composed of high-Tc cuprates and  manganite materials.
In our study, proximity effects are induced by the interplay of two separate interface effects:
spin-mixing (or rotation) surface scattering and spin-flip scattering.
The surface spin-mixing scattering introduces spin-triplet pairing correlations in
superconducting side; as a result, Andreev bound states are formed at energies within 
the superconducting gap. The spin-flip scattering
introduces not only long range equal-spin
pairing amplitudes in the half-metal,
but also an exotic magnetic proximity effect extending into the superconductor.
\end{abstract}

\maketitle


The possibility of making heterostructures out of
high-T$_{\rm c}$ cuprate superconductors and strong manganite ferromagnets
like LSMO allow us to study the competition between superconducting and
magnetic order where both are of equal strength.
The magnanites are of special interest since they have the same 
perovskite structure as the cuprates and e.g. L$_{0.67}$Sr$_{0.33}$MnO$_{3}$ 
is close to totally spin
polarized i.e. a half metal\cite{Park1998}. 
We report initial results from an on-going study of the electron density
of states (DoS) and the induced magnetism and superconductivity in
half-metal/$d$-wave superconductor/half-metal trilayers.
In particular we wish to identify properties that would
allow us to distinguish between a structure having the two half-metals
with parallel spin-bands from a one having them anti-parallel.

The study is based on quasiclassical Green's function theory and the
key issue is to pose the correct boundary conditions that connects
a superconducting half-space with a half-metallic one where only one spin-band
is present\cite{MRS88,Tokuyasu88,Fogelstrom2000,Cuevas2001,Eschrig2003}. 
In order to have coherent transport between a superconductor and a half-metal
the interface needs to be spin-active, i.e. to reflect or transmit quasiparticles 
(and quasiholes) in a manner that depends on the orientation of their spin relative
to the magnetic orientation that defines the spin-active interface. There are two
different ways of spin-active scattering the interface may possess that
we take into account: {\bf i)} {\it spin-mixing} or {\it spin-rotation},
the quasiparticle (quasihole) acquires a phase shift, $\pm \Theta_{m}/2$, with the sign depending
its spin orientation. The presence of spin-mixing introduces Andreev states in the DoS
below the gap-energy $\Delta$ and an (magnetic) exchange field in the superconductor
\cite{Tokuyasu88,Fogelstrom2000}.
{\bf ii)} {\it spin-flip scattering},
the half-metal allows transmission only of one
spin orientation which leads to two channels, with hopping amplitudes 
$\tau_{\uparrow\!\uparrow}$ and $\tau_{\downarrow\!\uparrow}$,
of spin-scattering \cite{Eschrig2003}. Through spin-flip scattering
equal-spin superconducting correlations can be admitted into the half-metal. 
These two ways of spin-activeness may be accounted for by the boundary conditions
the quasiclassical Greens function must obey at the superconductor/half-metal interface
via the set of {\it phenomenological} parameters
$(\Theta_{m},\tau_{\uparrow\!\uparrow},\tau_{\downarrow\!\uparrow})$ \cite{Eschrig2003}.

\begin{figure}[t]
  \includegraphics[width=1.0\textwidth]{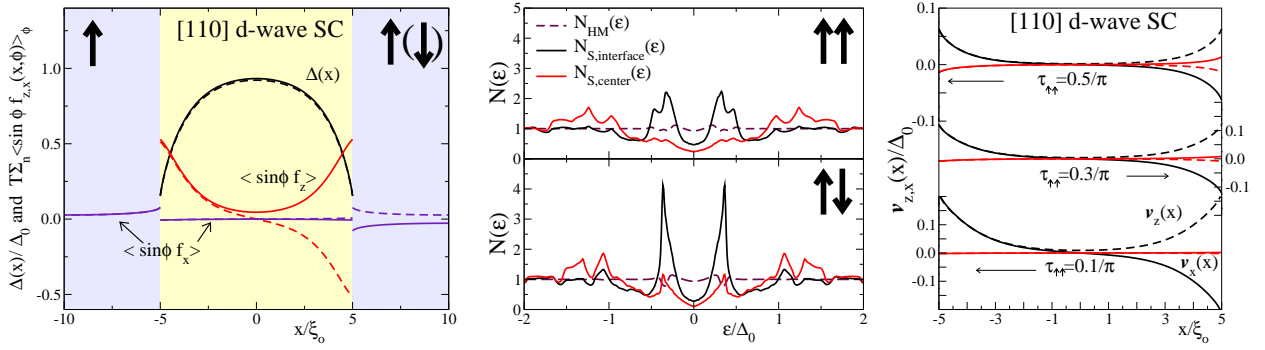}
  \caption{To the left we plot the amplitude of the d-wave orderparameter as well as the triplet
pairing correlations (TPC). The TPC arise from the (repeated) scattering off the spin-active 
interfaces and they are also induced as equal-spin TPC 
($T \sum_n \langle \sin \phi f_x(x,\phi;\epsilon_n)\rangle$) in the half-metal. 
The interface parameters are [$\Theta_m=\frac{\pi}{4},\tau_{\uparrow\!\uparrow}=\frac{0.1}{\pi},
\tau_{\downarrow\!\uparrow}=0.9\, \tau_{\uparrow\!\uparrow}$] which corresponds to a low-transmission
interface that is strongly spin-active. The full lines correspond to parallel alignment of the
conduction bands of the two half-metals and dashed lines to anti-parallel alignment. In the center
we show the total density of states (DoS) both in the half-metal and in the superconductor. 
For the superconductor we show the DoS at the interface and in the center (x=0). Finally to the
right we show the induced exchange field in the superconductor for three values of transparency 
of the interface. Exchange fields $\nu_{x,z}$ 
are calculated assuming a Fermi-liquid parameter $A^a_0=-0.7$. 
The full (dashed) lines corresponds to 
(anti-)parallel half-metal alignment.}
\label{Fig1}
\end{figure}

The geometry we study is that of two equal length half-metals, length $L_{HM}$, that
sandwich a d-wave superconductor, length $L_S$. Both types of material are assumed to conduct in
two-dimensional planes and the contact between them
is made between these planes. 
Furthermore we consider
either that the two half-metals are parallel, both conduct in the same spin band, or they
are anti-parallel and conduct in opposite spin bands. The sizes of the systems are taken
to be on the order of 10 $\xi_0$ ($=\hbar v_f/2\pi k_B T_c$, the coherence length
of the superconductor) and we assume clean materials so that impurity scattering may be neglected. 
Finally, since we consider a d-wave superconductor the
ab-plane of the superconductor may be oriented with respect to the interface normal. 
5A
It is well known that the d-wave superconductor admits zero-energy bound states (ZEBS) in
the DoS
if the ab-plane is not aligned to the interface \cite{Hu94}. These ZEBS will be shifted
from zero energy by a finite spin-mixing scattering and the DoS of 
the 45$^o$-degree rotated (or [110]) d-wave turns out to be very
sensitive to weak spin-mixing.

In Figure \ref{Fig1} we display the orderparameter profile in the d-wave superconductor, 
the triplet pairing 
correlations (TPC) both in the superconductor 
and in the half-metal, and the angle averaged
(total) DoS at various locations in our structure for a representative set of parameters.
The temperature is set to $0.1 {\rm T_c}$.
In addition to finite size effects, 
the orderparameter is strongly suppressed
in the vicinity of the interface due to 
the sign nature of d-wave pairing function. 
However, its magnitude at the interfaces is 
slightly enhanced by the spin-mixing scattering  
compared to that of a d-wave superconductor 
at a spin-inactive interface. 
The spin-mixing scattering generates $S_z=0$ 
TPC $(f_z)$ in the superconductor while
spin-flip scattering induces equal-spin $S_z=\pm1$ TPC 
both in half-metal and in the superconductor $(f_x)$. 
Although parallel or anti-parallel 
alignment hardly makes a difference on the magnitude 
of the orderparameter, 
the TPC in parallel and anti-parallel alignments
shows symmetric or antisymmetric behavior with 
respect to the center, respectively. 
For more transparent interfaces the equal-spin TPC grow stronger and
gives rise to an exchange field 
which direction-axis rotates in the superconductor.
Typical magnitude for the exchange fields is 
$\sim 10\%$ of $\Delta_0$ at the interface
and decaying into the center of the superconductor. Note also
that for the [110]-crystal the exchange field has the
opposite symmetry with regard to the center-point 
compared to the TPC. For the [100]-crystal these symmetries
are the same and the TPC have a $\cos\phi$ p-wave basis function
compared to the $\sin\phi$ for the [110]-crystal.

To investigate the alignment effects further, 
we show the DoS for both cases in the center panel of Figure \ref{Fig1}. 
For both half-metal alignments, the ZEBS split in to
two peaks in the DoS on the superconductor side owing to 
the spin-mixing scattering\cite{Fogelstrom2000}.  
There is also a change in the DoS in the half-metal sides 
via the proximity effect. This is due to the spin-flip scattering. 
We find the alignment difference clearly manifested
in the DoS in superconductor side. 
In contrast to the sharp peaks in the DoS the anti-parallel alignment shows, 
the bound-states peaks in parallel alignment are broadened or even split
by finite size effects. 
This difference is due to that
interference of repeated spin-mixing scattering
of the bound states in parallel alignment is constructive
and hence a split bound-state signature DoS. For the
anti-parallel alignment this constructive interference is
canceled due to that consecutive scattering picks up
opposite phase kicks at either interface.


We gratefully acknowledge financial support 
from "Grant-in-Aid for 
Scientific Research" from the MEXT of Japan (N.Y) 
and the Swedish Research Council (M.F). 



\bibliographystyle{aipproc}   


%
%

\end{document}